\documentclass[11pt,a4paper]{article}
\usepackage[T1]{fontenc}
\usepackage[margin=27mm]{geometry}
\usepackage{graphicx,amsmath,amssymb,booktabs,array}
\usepackage{url,placeins}

\usepackage[hidelinks]{hyperref}
\graphicspath{{figures/}}
\title{Shortest-Path Multiplicity Heterogeneity as a Graph Diagnostic for Radial Organization in Sampled Geometries}
\author{Ariadna Uxue Palomino Ylla\\Nagoya University, Nagoya, Japan\\\texttt{ariadnauxue@gmail.com}}
\date{}
\begin{document}
\maketitle
\begin{abstract}
Shortest-path counts describe how many minimal routes connect a pair of graph vertices, complementing distance alone. We study their heterogeneity within graph-distance shells using $C_{\log}(p,r_g)$, the cube root of the third absolute central moment of log shortest-path multiplicities. This permutation-invariant statistic measures variation across shell targets and does not encode direction by itself. In a Flamm--Schwarzschild case study, its radial profile differs from that of a matched-flat control. The binned statistic has a Pearson correlation of approximately $0.912$ with a logarithmic Schwarzschild reference profile, indicating radial-reference agreement rather than reconstruction of continuum curvature. Comparisons across sample sizes and neighbor parameters further describe behavior under the tested graph constructions. Preserved summaries and plotting records support the numerical evidence, although original graph realizations and the definitive execution state were not recovered. These results motivate multiplicity heterogeneity as a graph diagnostic for radial organization, with interpretation conditional on sampling, connectivity and boundary effects.

\end{abstract}

\section{Introduction}
Spatial networks arise when nodes are embedded in, sampled from, or constrained by a geometric domain. Connectivity and paths can retain information about that domain. They also reflect sampling density, finite boundaries and the rule used to connect points. The practical question is therefore which aspects of geometric organization a graph observable describes, and which alternative mechanisms can produce a similar pattern.

We study the variation of shortest-path multiplicities across a graph-distance shell. For a source $p$, each target at distance $r_g$ may be reached by one or several shortest paths. These counts describe the degeneracy of minimal routes, information not contained in the distance alone. Their logarithmic dispersion defines $C_{\log}(p,r_g)$. It can be computed from unweighted graph topology once the graph has been specified.

The earlier manuscript called this quantity shortest-path anisotropy. We use \emph{shortest-path multiplicity heterogeneity} to make its mathematical content explicit. Permuting shell targets while keeping their counts leaves $C_{\log}$ unchanged. Consequently, this scalar is not a directional statistic unless combined with additional information about target directions. Heterogeneity can arise from graph construction, sampling, bottlenecks or boundaries, as well as geometric organization. Homogeneity or isotropy alone does not supply a quantitative finite-graph null distribution for this observable.

The numerical case study concerns a Flamm--Schwarzschild spatial slice and a matched-flat control. We examine radial profiles, association with an external radial reference, comparisons across graph sizes, and sensitivity to the neighbor parameter. Section~\ref{sec:provenance} describes the preserved numerical evidence and limits on reproduction.

For the Flamm benchmark the external reference is the Schwarzschild Kretschmann scalar,
\begin{equation}
K_{\mathrm{Schw}}(r)=\frac{48M^2}{r^6}.
\end{equation}
This scalar belongs to the parent spacetime; it is not the intrinsic curvature of the sampled spatial surface. It supplies a simple associated radial profile. At fixed mass, $\log K_{\mathrm{Schw}}$ is an affine function of $\log r$, so strong correlation with it does not establish curvature specificity. The control comparison is informative about the graph constructions represented here, but cannot by itself remove sampling or construction confounding.

Our contribution is a simple shell-based dispersion measure together with an exploratory spatial-graph case study. The observed numerical associations are distinct from directional interpretation, continuum reconstruction and claims of generality. Quantitative extensions to other geometries and comparisons with simpler graph observables remain outside the evidence presented here.

\section{Related Work}
Spatial constraints influence connectivity and transport in networks, as reviewed by Barth\'elemy~\cite{Barthelemy2011}. Random geometric graphs provide a mathematical setting in which sampled positions and a proximity rule jointly determine edges~\cite{Penrose2003}. These perspectives motivate attention to the sampling measure, domain and distance rule when interpreting shortest paths. A graph statistic can respond to the construction as well as to the sampled geometry.

Graph-distance balls and shells organize neighborhoods beyond immediate adjacency. Costa and Silva describe hierarchical network measurements using successive distance rings and balls, including counts and connectivity within and between rings~\cite{CostaSilva2006}. The present observable uses the same distance-based organization, but summarizes shortest-path multiplicities across targets at a fixed distance. Shell size and shell connectivity are natural comparison quantities, although no comparative results are reported here.

Shortest-path multiplicity is an established ingredient of centrality analysis. Brandes' betweenness algorithm counts shortest paths while computing distances, then uses those counts to accumulate contributions to centrality~\cite{Brandes2001}. Our statistic instead takes the dispersion of log counts over a source's shell. It is neither a new counting algorithm nor a betweenness measure: the distinction lies in the aggregation of route degeneracy at fixed graph distance.

Graph curvature offers a different route from connectivity to geometric description. Forman's combinatorial construction and Ollivier's metric-measure approach define curvature through structures distinct from shell-target dispersion~\cite{Forman2003,Ollivier2009}. Ni and colleagues apply network Ricci flow to community detection~\cite{Ni2019}, illustrating a use of graph curvature beyond comparison with a continuum reference. These approaches provide future comparators; conceptual differences alone do not establish added explanatory value for $C_{\log}$.

Spatial baselines must account for mechanisms imposed by an embedding. In community detection, Expert and colleagues incorporate spatial effects into a null model to distinguish them from other network organization~\cite{Expert2011}. Their application differs from the present problem, but motivates testing which spatial constraints a control preserves. Here, a matched-flat graph is an initial control, not a complete null ensemble: matching node count and nominal $k$ need not match sampling measure, shell size, local edge scale or boundary exposure.

\section{Method}
\subsection{Spatial graphs and construction dependence}
Let $\mathcal{M}$ be a geometric domain and let sampled points form the vertices of an unweighted, undirected graph $G=(V,E)$. A proximity rule determines its edges. Once $G$ is fixed, $d_G(p,q)$ denotes the number of edges in a shortest path. The observable below does not require embedding coordinates; coordinates enter graph construction and subsequent radial analysis.

A $k$-nearest-neighbor relation is directed before symmetrization. An undirected union rule connects a pair whenever either point selects the other. A radius rule instead connects pairs within a prescribed distance. These rules are not interchangeable at fixed $k$. The recovered sources contain both constructions and conflicting sampling definitions. Section~\ref{sec:construction-evidence} distinguishes these sources rather than assigning an unverified single protocol to every result.

\subsection{Graph shells and shortest-path multiplicity}
For a node $p\in V$ and an integer graph radius $r_g\geq 1$, define the graph shell
\begin{equation}
S_{r_g}(p)=\{q\in V: d_G(p,q)=r_g\}.
\end{equation}
The corresponding graph ball is
\begin{equation}
B_{r_g}(p)=\{q\in V: d_G(p,q)\leq r_g\}.
\end{equation}
For every target node $q\in S_{r_g}(p)$, let
\begin{equation}
N_{\mathrm{geo}}(p,q)
\end{equation}
be the number of shortest paths from $p$ to $q$. It is at least one for a reachable shell target. Unreachable vertices do not belong to a finite-distance shell. Distance specifies the length of a minimal route, whereas multiplicity counts how many such routes exist.

\begin{figure}[!htbp]
\centering
\includegraphics[width=\textwidth]{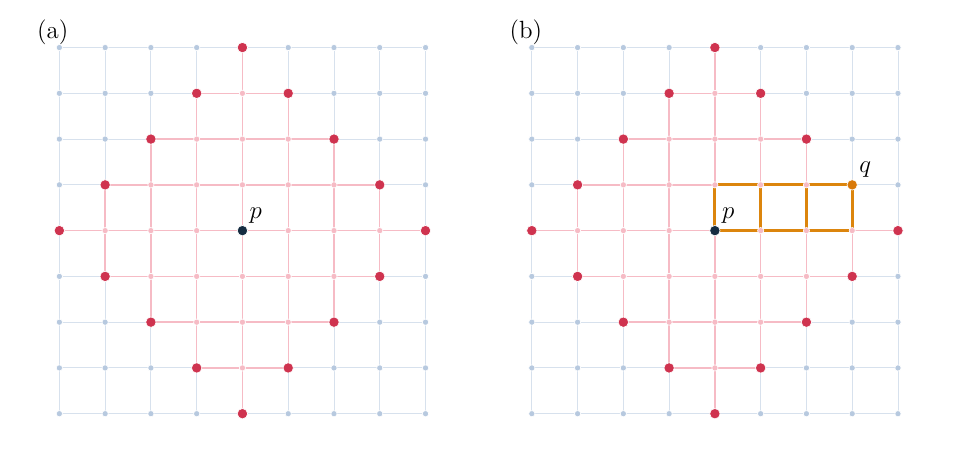}
\caption{Schematic illustration on a square lattice. (a) The ball $B_{r_g}(p)=\{v:d_G(p,v)\leq r_g\}$ contains the source $p$ (dark) and interior nodes (light pink); the shell $S_{r_g}(p)=\{v:d_G(p,v)=r_g\}$ is red, with $r_g=4$. Blue nodes lie outside the ball. Here $d_G$ counts edges in a shortest path. (b) Orange edges show the union of the four shortest paths to the orange target $q$, so $N_{\mathrm{geo}}(p,q)=4$. The statistic uses multiplicities for all shell targets. This graph is illustrative, not experimental data.}
\label{fig:shell-schematic}
\end{figure}

\subsection{Logarithmic multiplicity heterogeneity}
For a fixed source node $p$ and graph radius $r_g$, we collect the logarithmic shortest-path multiplicities
\begin{equation}
X_{p,r_g}=\{\log N_{\mathrm{geo}}(p,q): q\in S_{r_g}(p)\}.
\end{equation}
The collection is indexed by shell targets, so repeated count values remain represented. We use the natural logarithm. It compresses large counts and reduces their dominance relative to an untransformed count scale; no cross-size stability theorem is asserted.

The observable is the cube root of the third absolute central moment of this collection:
\begin{equation}
C_{\log}(p,r_g)=
\left[\frac{1}{|S_{r_g}(p)|}\sum_{q\in S_{r_g}(p)}
\left|\log N_{\mathrm{geo}}(p,q)-\mu_{p,r_g}\right|^3
\right]^{1/3},
\label{eq:clog}
\end{equation}
where
\begin{equation}
\mu_{p,r_g}=\frac{1}{|S_{r_g}(p)|}\sum_{q\in S_{r_g}(p)}\log N_{\mathrm{geo}}(p,q).
\end{equation}
This is an absolute central-moment dispersion, not a signed third moment or a skewness coefficient. The cubic power emphasizes larger departures from the shell mean while retaining a simple definition. Its use is a methodological choice, not a claim of unique optimality or an established advantage over the standard deviation of log counts.

The statistic is invariant under every permutation of the shell targets. It measures multiplicity heterogeneity, not the angular arrangement of high- and low-count targets. Directional analysis would require additional target-position information. Equal multiplicities give zero dispersion, while unequal multiplicities give positive dispersion; neither case identifies the underlying geometric cause.

Recovered notebook implementations exclude shells with at most two targets and evaluate logarithms with a small additive offset $10^{-12}$. The mathematical definition uses positive integer path counts without that offset. Packaged variants differ, as discussed below. Small-shell exclusion is a resolution criterion; it does not generally remove all boundary-affected sources or paths.

\subsection{Radial aggregation}
A node inherits a radial coordinate $r(p)$ from the sampled domain. For a radial bin $\mathcal{B}_i$ of admissible source nodes, the mean statistic is
\begin{equation}
\overline{C}_{\log}(r_i,r_g)=\frac{1}{|\mathcal{B}_i|}\sum_{p\in\mathcal{B}_i}C_{\log}(p,r_g).
\end{equation}
The coordinate used for correlation is the mean source radius in a bin, except where the common-bin routine pools the two geometries. Binning details are result-specific. A logarithm of a bin mean of $K$ is not generally equal to the mean of $\log K$, or to $\log K$ evaluated at mean radius.

\begin{figure}[!htbp]
\centering
\includegraphics[width=\textwidth]{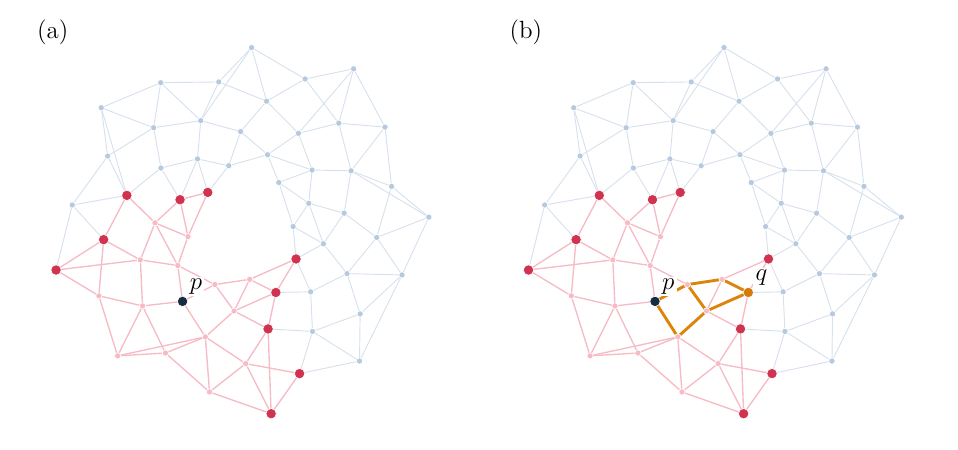}
\caption{Schematic illustration on a deterministic irregular graph embedded on a Flamm surface, shown in $x$--$y$ projection. (a) The source $p$ is dark, interior ball nodes are light pink, shell nodes are red and outside nodes are blue. The ball is $B_{r_g}(p)=\{v:d_G(p,v)\leq r_g\}$ and the shell is $S_{r_g}(p)=\{v:d_G(p,v)=r_g\}$, with $r_g=3$; $d_G$ counts edges, not drawing distance. (b) Orange edges show all shortest paths to the orange target $q$: $d_G(p,q)=3$ and $N_{\mathrm{geo}}(p,q)=3$. Multiplicity heterogeneity uses all shell targets. This illustrative graph is separate from the numerical benchmark.}
\label{fig:curved-shell-schematic}
\end{figure}
\FloatBarrier

\section{Experimental Design and Provenance}
\subsection{Scope of the benchmark}
The quantitative evidence is restricted to the Flamm--Schwarzschild versus matched-flat case study. The surface representation in the recovered notebook is $z=2\sqrt{2M(r-2M)}$, with $M=1/2$. Numerical coordinates use that mass/length convention; no physical-unit calibration or intrinsic-geodesic nearest-neighbor construction is established by the recovered evidence.

The matched-flat idea is to remove embedding height while retaining planar radial and angular organization. Matching these coordinates and node count does not imply matching intrinsic area measure, local edge scale, shell size, degree or exposure to boundaries. Construction differences make these distinctions especially important. Hyperbolic, Reissner--Nordstr\"om, Bardeen and Hayward extensions are not treated as quantitatively validated benchmarks here.

\subsection{Recovered, inferred and unresolved construction details}
\label{sec:construction-evidence}
\paragraph{Recovered source definitions.}
Early notebook definitions sample the angle uniformly on $[0,2\pi]$, followed by a radius uniform on $[1.05,5]$, and use $M=1/2$. This is uniform radial sampling, not uniform intrinsic surface area. A recovered Flamm builder uses an ambient Euclidean distance matrix in three dimensions and a calibrated radius equal to $1.15$ times the median distance to the $k$th nearest neighbor; pairs within that radius are connected. Here $k$ calibrates a radius rather than specifying a fixed degree.

Historical package definitions instead include annular sampling with radius equal to the square root of a uniform draw in squared radius and a lower boundary $2M+0.2$, equal to $1.2$ here. That convention is area-uniform in the planar annulus, not necessarily in the curved surface's intrinsic area. A pre-PDF package defines the matched-flat helper by removing height and forming an undirected union $k$-nearest-neighbor graph. Ambient Euclidean proximity should not be described as intrinsic geodesic proximity.

\paragraph{Inferred links.}
Saved export filenames, input statements and plotted values link the notebook to Figures~\ref{fig:radial-profiles}--\ref{fig:robustness}. These are evidence for candidate generation routes. An earlier-dated package or matching seed literal is not proof that its definitions were active when a saved result was computed.

\paragraph{Unresolved execution details.}
The notebook and package versions differ in sampling, boundaries, graph rules and estimator details. Original graph realizations and the definitive execution state were not recovered, so no unique historical protocol can be certified for all results. Identical curved and flat edge rules are not established. The final Data and provenance statement details the execution evidence underlying this uncertainty.

\subsection{Parameters, binning and statistical units}
Table~\ref{tab:protocol} distinguishes the analyses. Entries are recovered from saved inputs and summary metadata; they do not resolve the active-builder ambiguity. All listed analyses use $M=1/2$ and $r_g=3$. Broader radius exploration in archival material is not substituted for the evidence shown here.

\begin{table}[!htbp]
\centering\small
\caption{Result-specific recovered settings. Candidate denotes a single-realization linkage from saved notebook statements rather than recovered graph state. These rows do not describe a full factorial experiment.}
\label{tab:protocol}
\begin{tabular}{>{\raggedright\arraybackslash}p{0.27\textwidth}>{\raggedright\arraybackslash}p{0.18\textwidth}>{\raggedright\arraybackslash}p{0.16\textwidth}>{\raggedright\arraybackslash}p{0.24\textwidth}}
\toprule
Analysis & $N$; $k$ & Seeds & Aggregation \\
\midrule
Text radial correlations & $1000$; $16$ & $1$--$5$ & 12 bins per profile \\
Figure 3, candidate & $1000$; $16$ & $1234$ & 12 common bins \\
Figure 4, candidate & $500$; $14$ & $1234$ & 12 plotted bins \\
Table 2 & $200$; $12$ & $1$--$10$ & Node-level ranks \\
 & $500$; $14$ & $1$--$10$ & Node-level ranks \\
 & $1000$; $16$ & $1$--$5$ & Node-level ranks \\
Figure 5 & $1000$; $12,14,16,18,20$ & $1$--$3$ per $k$ & 12 bins per profile \\
\bottomrule
\end{tabular}
\end{table}

For Figure~\ref{fig:radial-profiles}, the common-bin routine divides the overlap of valid radial ranges into 12 equal-width bins. It includes the upper endpoint by clamping the bin index, retains bins represented in both geometries, and uses their pooled mean radius for plotting. Counts and means are computed within each geometry. Saved plot coordinates survive, but original node-to-bin assignments and complete bin-count data do not.

For Figure~\ref{fig:logK}, the recovered routine uses 12 equal-width radial bins with lower-inclusive, upper-strict intervals and omits bins with fewer than five admissible rows. Its horizontal coordinate is $\log\overline{K}_i$, where $\overline{K}_i$ is the mean reference value in bin $i$. The saved input specifies seed $1234$; its generation sequence is documented in the final Data and provenance statement.

The five-seed text estimates and Figure~\ref{fig:robustness} instead use the recovered separate-profile radial-binning routine. Pearson correlations are computed across radial-bin mean radius and mean $C_{\log}$, separately for each realization and geometry. The reported error magnitude is the sample standard deviation of those realization-level coefficients, neither a standard error nor a confidence interval. Curved and flat outputs are paired within each seed; the same seed labels recur across $k$.

Table~\ref{tab:refinement} concerns node-level Spearman rank correlation between reference values and $C_{\log}$. The recovered rank routine assigns average ranks to ties. Each seed coefficient was formatted to three decimal places before being converted back to a number and averaged. The listed means retain that historical rounding procedure. They have been checked against saved summaries, not recomputed from complete original per-node data.

\subsection{Reproducibility and provenance}
\label{sec:provenance}
The evidence consists of preserved manuscript source, numerical summaries and plotting records, together with deterministic replacement schematics. Original graph realizations and the definitive execution state were not recovered. Verification concerns the reported summaries and plotted coordinates, not exact end-to-end reproduction of the sampled graphs.

Figures~\ref{fig:shell-schematic} and~\ref{fig:curved-shell-schematic} are reproducible schematic replacements, separate from the numerical experiments. Figures~\ref{fig:radial-profiles}--\ref{fig:robustness} present the archived numerical plots; their plotted data, axes and error bars are preserved. The historical axis label LogCMD refers to multiplicity heterogeneity. Saved seed-level records support the five-seed radial estimates and Figure~\ref{fig:robustness} summaries. Saved plotting coordinates support the Figure~\ref{fig:logK} coefficient, and notebook summary output supports Table~\ref{tab:refinement}. A conflicting Figure~\ref{fig:logK} tabular file remains unresolved, as detailed in the final Data and provenance statement. No new simulation or baseline experiment is included.

\section{Results}
\subsection{Radial organization of multiplicity heterogeneity}
The Flamm profile decreases with radius, whereas the matched-flat profiles are weaker and more variable across seeds. The recovered five-seed summaries give
\begin{equation}
\mathrm{Corr}(r,\overline{C}_{\log})_{\mathrm{Flamm}}=-0.960\pm 0.015,
\end{equation}
and
\begin{equation}
\mathrm{Corr}(r,\overline{C}_{\log})_{\mathrm{MatchedFlat}}=0.180\pm 0.315.
\end{equation}
Here the notation denotes the mean and sample standard deviation of five realization-level Pearson coefficients, with the settings in Table~\ref{tab:protocol}. It does not denote a node-level correlation or uncertainty of the mean. Values are reported at their original precision.

\begin{figure}[!htbp]
\hspace{2.01cm}
\includegraphics[width=0.85\textwidth]{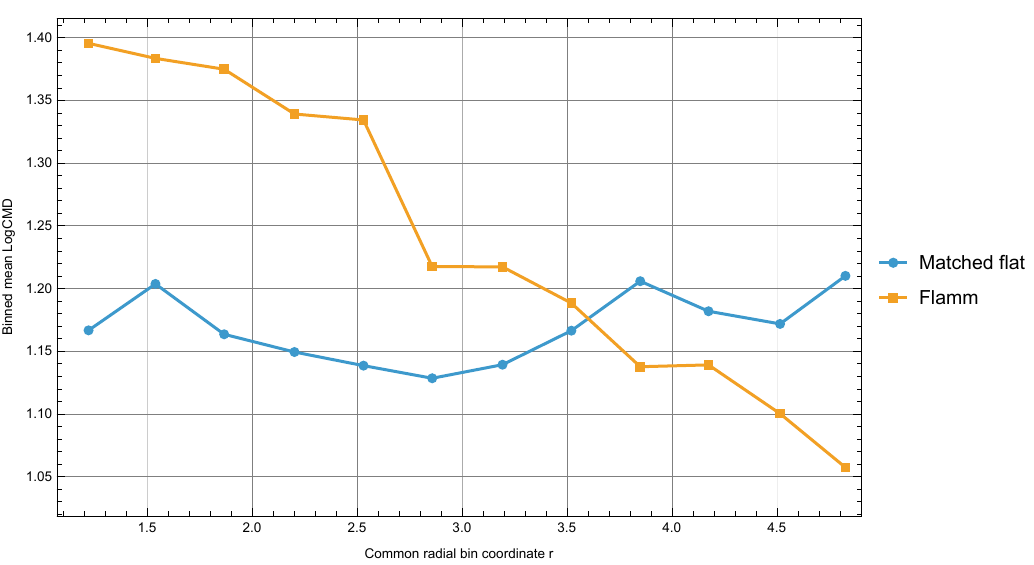}
\caption{Mean multiplicity heterogeneity in the Flamm graph and matched-flat control on common radial bins for one realization. This comparison is separate from the five-seed correlation estimates in the text.}
\label{fig:radial-profiles}
\end{figure}

Figure~\ref{fig:radial-profiles} shows a separate single-realization comparison using common bins. It is not a visualization of the five-seed means above. The contrast describes radial organization under the represented constructions; it does not establish that sampling, boundary effects or edge-rule differences have been excluded.

\subsection{Agreement with an external radial reference}
The recovered plotting coordinates give
\begin{equation}
\mathrm{Corr}\left(\log\overline{K}_i,\overline{C}_{\log,i}\right)\approx 0.912.
\end{equation}
This is a Pearson correlation across the 12 saved plotted bins, with $\overline{K}_i$ defined by the recovered binning routine. The plotted-point coefficient agrees with the reported value.

The Kretschmann profile is used because it is a known radial function associated with the parent Schwarzschild spacetime of the spatial benchmark. The graph samples a spatial surface, so this is an external reference, not its intrinsic curvature. At fixed $M$, $\log K_{\mathrm{Schw}}(r)=\log(48M^2)-6\log r$. A statistic varying smoothly with radius can correlate strongly with this reference without specifically reconstructing curvature. Bin averaging changes the precise coordinate transformation but does not eliminate this limitation.
\begin{figure}[!htbp]
\centering
\includegraphics[width=0.70\textwidth]{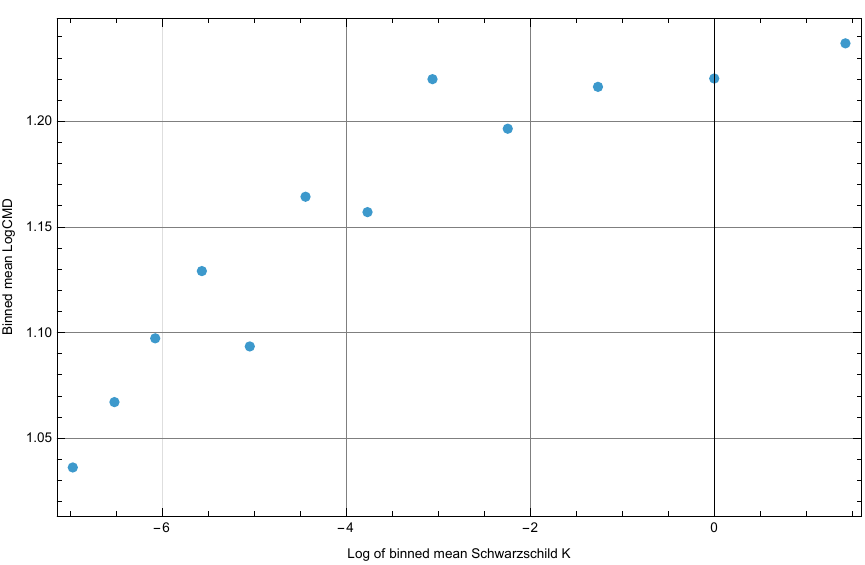}
\caption{Binned multiplicity heterogeneity versus the logarithm of the bin-mean Schwarzschild reference. The 12 plotted points give Pearson correlation approximately $0.912$, describing radial-reference agreement.}
\label{fig:logK}
\end{figure}

\subsection{Comparison across sample sizes}
Table~\ref{tab:refinement} reports the three mean Spearman values. The means increase across these configurations, but $k$ and the number of seeds also change with $N$. This is not an isolated refinement experiment and does not establish convergence, improved stability or separation from finite-sample noise.
\begin{table}[!htbp]
\centering
\caption{Mean node-level Spearman agreement between the reference and graph observable in the Flamm benchmark, supported by saved summary output. Configurations differ in $k$ and seed count (Table~\ref{tab:protocol}); seed coefficients were rounded before averaging.}
\label{tab:refinement}
\begin{tabular}{cc}
\toprule
Number of sampled nodes & Mean Spearman agreement \\
\midrule
$N=200$ & $0.5320$ \\
$N=500$ & $0.5630$ \\
$N=1000$ & $0.6636$ \\
\bottomrule
\end{tabular}
\end{table}
\FloatBarrier
These values provide a descriptive comparison across configurations. A controlled resolution analysis would require a documented construction, suitable replication and explicit treatment of changes in connectivity scale. No such additional experiment is reported here.

\subsection{Sensitivity to the neighbor parameter at fixed shell radius}
\begin{figure}[!htbp]
\hspace{2.01cm}
\includegraphics[width=0.8\textwidth]{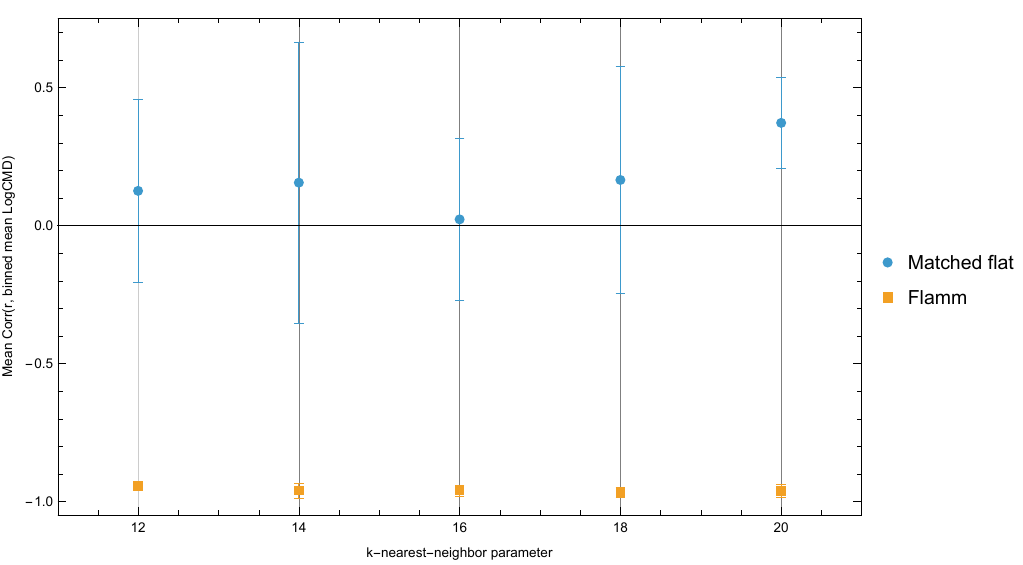}
\caption{Radial Pearson correlation versus $k$ at $N=1000$ and fixed $r_g=3$. Points are means and error bars are plus or minus one sample standard deviation across three seeded realizations (seeds $1$--$3$), with $M=1/2$ and 12 bins per profile. Curved and flat outputs are paired by seed. Bars do not represent shell-radius variation or uncertainty of the mean.}
\label{fig:robustness}
\end{figure}

Figure~\ref{fig:robustness} presents the $k$ scan at $N=1000$ and fixed $r_g=3$. Flamm correlations remain strongly negative across the displayed $k$ values, while matched-flat means are weaker and their variability is larger. This is a descriptive statement about the displayed scan, not general robustness across graph radii or construction protocols.

Error bars are sample standard deviations of realization-level Pearson coefficients across three seeds at each $k$, with $r_g=3$ fixed. They measure between-realization variation, not variation across shell radii or uncertainty of the mean. The three-seed $k=16$ point is distinct from the five-seed estimates quoted above.

\FloatBarrier

\section{Discussion}
The results show that shortest-path multiplicity heterogeneity can exhibit pronounced radial organization in a sampled spatial graph. This is the positive empirical observation of the study. The scalar is simple to define on unweighted topology, and comparison with a flat construction provides an initial reference for interpreting its radial pattern. The evidence does not establish a directional observable, a universal curvature measure or curvature-specific discrimination.

Graph construction is central. Ambient distances, intrinsic distances, union $k$-nearest-neighbor relations and calibrated-radius graphs can produce different shortest-path degeneracies from related point sets. A comparison at equal node count and nominal $k$ does not necessarily match degree, local edge scale or shell size. The recovered version conflicts prevent definitive attribution of the observed contrast to geometry alone.

Sampling and boundaries provide further alternatives. Uniform sampling in radius, planar area and intrinsic area are different measures. Even with shared radial and angular coordinates, curved and flat embeddings need not have comparable local spatial density or connectivity. Excluding small shells does not guarantee that retained shells and their shortest paths avoid a boundary. Radial occupancy, distances to boundaries and local edge scales would need direct examination to strengthen the matched-control interpretation.

The reference comparison must also remain limited in scope. A Schwarzschild spacetime scalar is not the intrinsic curvature of the sampled surface. Correlation with its monotonic radial profile records association with radius-dependent organization, rather than a reconstruction theorem or demonstration of curvature specificity. An intrinsic-spatial reference comparison and controlled null ensembles would address different questions from the present plot.

Provenance limits are substantive. Saved summaries and plotting coordinates support retained numerical reporting, but missing original graph realizations and unresolved active code definitions restrict verification of the generation process. The pedagogical schematics provide no evidence about the missing numerical execution history. This distinction matters when assessing the method's current empirical support.

Several comparisons remain future work: standard deviation of log multiplicities, shell size, degree, clustering, radial occupancy, local edge scale, and established graph-curvature measures. Homogeneous isotropic spatial ensembles under matched construction, and additional network null models, could quantify expected fluctuations in the absence of the proposed signal. These comparisons have not been established by the available evidence. Directional analysis, weighted graphs, and hyperbolic or other black-hole-inspired benchmarks likewise require separately documented validation.

\section{Conclusion}
We have presented shell-based shortest-path multiplicity heterogeneity as a graph observable and examined a Flamm versus matched-flat case study. The core statistic is a cubic absolute central-moment dispersion of log path counts. It captures variation in route multiplicity at fixed graph distance without encoding target directions.

The results show radial organization and association with the external Schwarzschild reference profile under the tested constructions. They motivate further study of the observable as a promising graph diagnostic, while leaving curvature specificity, controlled baseline comparisons and exact graph-level reproduction unresolved.

\section*{Data and provenance statement}
This manuscript draws on preserved source, plotting evidence and numerical summaries documented in the accompanying provenance records. The claim-to-file mapping is recorded in \path{CLAIMS_AND_PROVENANCE_TABLE.md}; original recovered files remain separate in \path{computational_provenance_frozen/}, with investigation records in \path{provenance_repair/}. The accompanying repository for the archived independent draft is available at \url{https://github.com/Uxuee/shortest-path-multiplicity-heterogeneity/tree/independent_draft_public_polish_oct2026}. Deterministic generators and illustrative graph data for the replacement schematics are in \path{reproducible_schematics/}. The clean vector artwork is rendered from those data by \path{independent_manuscript/figure_design/build_clean_figures.py}. These schematics are pedagogical illustrations and are not experimental realizations.

The source notebook contains repeated definitions and nonsequential evaluation labels. Its original kernel state is unavailable. The matched-flat helper invoked by some scans lacks a definition in the extracted notebook inputs, although a packaged implementation exists. That implementation cannot be certified as the one active at execution. Original point clouds, complete graphs, raw per-source path data and complete node-to-bin assignments have not been recovered.

For Figure~\ref{fig:logK}, the saved input resets the seed to $1234$ and generates flat, hyperbolic and Flamm datasets in sequence; preceding random draws therefore matter to reproduction. This sequence does not establish quantitative validation of the hyperbolic benchmark. A related tabular file disagrees with the saved plot in its final bin and lacks the mean-reference column needed for the horizontal coordinate. The discrepancy remains unexplained. The figure and reported coefficient follow the saved plotting coordinates, without substituting the inconsistent table. Seed labels and candidate source versions do not resolve these execution and data gaps, and exact end-to-end reproducibility is not claimed.

\begingroup
\raggedright

\endgroup
\end{document}